\documentclass[prl,preprintnumbers,twocolumn,aps,superscriptaddress]{revtex4}
\usepackage{graphicx,}
\usepackage{graphicx,dcolumn,bm,epsfig,amsmath,amssymb,}
\usepackage{float}

\begin{document}

\title {Formation of topological vortices during superfluid transition
in a rotating vessel}

\author{Shreyansh S. Dave}
\email{shreyansh@iopb.res.in}
\author{Ajit M. Srivastava}
\email{ajit@iopb.res.in}
\affiliation{Institute of Physics, Bhubaneswar 751005, India}
\affiliation{Homi Bhabha National Institute, Training School Complex,
Anushakti Nagar, Mumbai 400085, India}


\begin{abstract}
 Formation of topological defects during symmetry breaking phase transitions
via the {\it Kibble mechanism} is extensively used in systems ranging from 
condensed matter physics  to the early stages of the universe. Kibble mechanism
uses topological arguments and predicts equal probabilities for the formation
of defects and anti-defects. Certain situations, however, require a net bias
in the production of defects (or antidefects) during the transition, for
example, superfluid transition in a rotating vessel, or flux tubes formation
in a superconducting transition in the presence of external magnetic field.
In this paper we present a modified Kibble mechanism for a specific system, 
$^4$He superfluid transition in a rotating vessel, which can produce the 
required bias of 
vortices over antivortices.  Our results make distinctive predictions which
can be tested in superfluid $^4$He experiments. These results also have 
important implications for superfluid phase transitions in rotating neutron 
stars and also for any superfluid phases of QCD arising in the non-central 
low energy heavy-ion collision experiment due to an overall rotation.

\end{abstract}

\maketitle

\noindent 

 Topological defects arise in a wide range of systems ranging from
condensed matter physics to the early stages of the universe.
Formation of these defects during symmetry breaking transitions has been
a very active area of research, especially in last few decades, bringing
out important interconnections between condensed matter physics and particle
physics. Indeed, the first detailed theory of formation of topological 
defects via a domain structure arising during a phase transition was
proposed by Kibble \cite{kbl} in the context of early universe. 
It was proposed by Zurek that certain aspects of Kibble mechanism
can be tested in superfluid helium systems \cite{zrk1}. It is now well
recognized that the basic physical picture of the Kibble mechanism applies 
equally well to any symmetry breaking transition \cite{zrk2,rjnt} thereby 
providing the possibility of testing the predictions of Kibble mechanism in 
various condensed matter systems, see refs.  \cite{he,sccor,lc1,lc2,lc3}. It 
is particularly important to note that the basic mechanism has many universal 
predictions making it possible to use condensed matter experiments to 
carry out rigorous experimental tests of these predictions made for cosmic
defects \cite{lc2,lc3}.  Defect formation in continuous transitions raises
important issues due to critical slowing down. The Kibble-Zurek mechanism
incorporates these aspects and leads to specific predictions of the
dependence of  defect densities on the rate of transition 
etc. \cite{zrk1,zrk2}.

 Basic physics of Kibble mechanism lies in the formation of a domain
structure during a phase transition where order parameter field 
varies randomly from one domain to another. Individual domains represent
correlation regions where order parameter field is taken to be uniform.
Another important physical input in the Kibble mechanism is the assumption
of smallest variation of the order parameter field in between the two
adjacent domains (the so called geodesic rule). With these two physical inputs,
a geometrical picture emerges for the physical region undergoing phase
transition, and straightforward topological arguments can be used to 
calculate the probability of formation of defects and anti-defects.
It is important to note that the probability of defect formation in the
Kibble mechanism is calculated {\it per correlation domain} and it is 
a universal prediction. Indeed, utilizing this universality, defect
formation probability for Kibble mechanism was experimentally tested in 
liquid crystal experiments \cite{lc1} for a first order transition case
where correlation domains could be directly identified as bubbles of the
nematic phase nucleating in the background of isotropic phase. However,
for a continuous transition, direct identification of correlation
domains is not possible. Further, here effects of critical slowing down
introduce dependence of relevant correlation length on the rate of
transition \cite{zrk2}. The Kibble-Zurek mechanism incorporates these
non-trivial aspects of phase transition dynamics for the case of continuous
phase transitions in prediction of defect density \cite{zrk1,zrk2}.
We now note that for the cases under consideration,
these topological calculations give equal  probability for the formation
of defects and anti-defects. Of course this is on the average, and
there can be excess of defects or antidefects in a given event of phase
transition. Kibble mechanism leads to important predictions about the typical 
value of this excess which, for the case of U(1) vortices in 2 space 
dimensions is found to be proportional to $N^{1/4}$ where $N$ is the total 
number of defects plus antidefects \cite{lc3}.

There are many physical situations which require a net excess of defects
or anti-defects (i.e. a non-zero value of the average net defect number)
in a phase transition due to external conditions.
For example, formation of flux tubes in type II superconductors in the 
presence of external magnetic field will lead to a net excess of
flux tubes oriented along the direction of external field.
Similarly, a $^4$He system undergoing a superfluid transition in a 
rotating vessel will lead to a net excess of vortices. 
Along with these excess defects (or anti-defects), there will also be
a random network of defects/antidefects resulting from domain structure
via the conventional Kibble mechanism. Normally, the net defect formation
(e.g. superfluid vortex formation in a rotating vessel) is studied
using arguments of energetics \cite{landau,hess1}. But the formation of 
superfluid vortices in a rotating vessel during the superfluid transition 
also includes contribution from a non-equilibrium defect production 
process (via the Kibble mechanism)  due to which number of formed vortices 
during the transition can deviate from the vortex model prediction.
(A deviation from the vortex model prediction was indeed observed
by Hess and Fairbank in their experiment \cite{hess2}, and in view of 
the above discussion, Kibble vortices may be able to account for this). 
As we elaborate below, in the presence of external influence (rotation of
initial fluid here, or external field for superconductor) the basic
physics of Kibble mechanism needs to be modified. 

Two most important ingredients of Kibble mechanism are, existence of 
correlation domains inside which the order parameter is taken to be
uniform, while the order parameter varies randomly from one domain 
to another, and the geodesic rule which
says that the order parameter variation in between two domains is
along the shortest path in the order parameter space. (We mention that
the geodesic rule becomes ambiguous for the case of superconductors
as discussed in \cite{gauge}. This makes our considerations of the present
paper non-trivial for superconductors, we will present it in a follow
up work.) We will show below that to get a net excess of defects or
antidefects in the presence of external influence (e.g. rotating vessel)
both of these aspects of Kibble mechanism need to be modified; a given
domain can no longer represent uniform value of the order parameter, 
rather each domain will have certain systematic variation of the order
parameter field originating from the external influence. Further,
the same external influence also affects the geodesic rule. In certain
situations, the variation of order parameter in between two adjacent domains
may trace a longer path on the vacuum manifold in apparent violation of
the geodesic rule.  We will show that this modified Kibble mechanism
leads to reasonable predictions of a net excess of defects, along with a
random network of defects/antidefects. Interestingly, it shows very
systematic deviations for the random component of the excess of defects
or antidefects from the Kibble prediction of $N^{1/4}$. We find that
this excess becomes larger with larger external bias. This is an important
prediction of the {\it biased Kibble mechanism} proposed here, and can be 
tested in experiments. This fluctuation in the net excess of defects resulting
from the phase transition, on top of the average net defect number 
arising from the rotation may account for the experimental results 
of Hess and Fairbank \cite{hess2} for superfluid transition in a rotating 
vessel where deviations from the energetics based net vortex number
(at times even negative vortex number) were found.

Superfluid component is characterized by a multi-particle condensate 
wave function, $\Psi=\Psi_0 e^{i\theta}$, where $\Psi_0^2$ gives 
number density of superfluid component. The superfluid velocity is
given by $\vec{v_s}=\frac{\hbar}{m}\vec{\nabla} \theta$,
where $m$ is the mass of $^4$He atom. We use the expression for the
free energy of the superfluid system in the presence of rotation 
\cite{landau,tilley} as $ F' = F-\vec{L}.\vec{\Omega}$,
where F is the free energy for superfluid without rotation and
$\vec{L}=\rho_s \int_{}^{} (\vec{r}\times \vec{v_s})d^2x$ is the angular 
momentum of the superfluid just after the phase transition generated due 
to external rotation ($\rho_s = m\Psi_0^2$  is the mass density), 
$\vec{\Omega}$ being the 
angular velocity of the vessel containing superfluid. Here we are assuming that 
part of normal component which undergoes superfluid condensation carries
same angular momentum as before the transition. (Though, it may be 
possible that only a fraction of the momentum of the normal fluid part 
which is condensing is carried over to the superfluid momentum. Effects of 
this possibility on our analysis requires a further study. One can determine
the value of this fraction experimentally using a rotating annulus of
the kind suggested in ref.\cite{zrk1}.) In two 
spatial dimensions, free energy density is given by,  
\begin{equation}
 f' = f -\rho_s (\vec{r}\times \vec{v_s}).\vec{\Omega} ,
\end{equation}
where $f$ is the free energy density of superfluid without any rotation. 
We thus get \cite{zrk2},
\begin{equation}
 f' = \alpha|\Psi|^2+\frac{\beta}{2}|\Psi|^4 + 
 \frac{\hbar^2}{2m}\Psi_0^2 |\vec{\nabla} \theta|^2 -
 \Omega \rho_s r \frac{\hbar}{m}|\vec{\nabla} \theta| ,
\end{equation}

\noindent where $\alpha$ and $\beta$ are phenomenological coefficients.
For temperatures less than the superfluid transition temperature, $\alpha
< 0$ and we determine the local value of condensate density $\Psi_0$ by
minimizing the free energy neglecting the rotation. (One can discuss
the effect of rotation on $\Psi_0$, even far away from vortices, 
especially in presence of boundaries. We keep analysis of this issue for 
future discussions.) With constant superfluid density $\Psi_0$, we minimize 
this free energy density with respect to $|\vec{\nabla}\theta|$ and get,
\begin{equation}
|\vec{\nabla} \theta|_{bias} = \frac{m\Omega r}{\hbar}.
\label{eq:bias}
\end{equation}
This shows that the equilibrium configuration of $\Psi$ requires  a non-zero 
value of $|\vec{\nabla} \theta|$ in the presence of rotation. (Note, for
the non-rotating case, we get $\theta = $ constant, as is assumed inside a 
domain in the conventional Kibble mechanism.) Note that  $|\vec{\nabla} 
\theta|_{bias}$ is proportional to the distance from the origin,
this will play an important role for the biasing in the production of 
vortices over antivortices as we will see below.

One of the main ingredients of Kibble mechanism is the randomness of the 
condensate phase $\theta$ from one correlated domain to other. As we
have discussed, for superfluid phase transition in the presence of rotation, 
order parameter $\theta$ cannot be uniform inside any domain, it
must vary systematically inside each domain.  
In this modified domain picture we still use the fact that all 
domains are independent from each other and have completely random $\theta$ 
value at the center of domain. (This type of picture was invoked in an 
earlier work by some of us where biased Skyrmion production due to
non-zero baryon chemical potential was studied via a modified Kibble 
mechanism for a toy model in 1+1 dimensions \cite{skyrmi}.) Further, 
the order parameter variation inside domain has to be such that 
it preserve the curl free motion of superfluid. 
As we have mentioned, here we are assuming that 
part of normal components which undergoes superfluid condensation carries 
the same angular momentum as before the transition, and we know that 
normal components follow rigid-body rotation with velocity given by
$\vec{v_n} = \Omega r \hat{\theta}$ which has non-zero curl. With transition 
to the superfluid phase, we model the domain structure in the presence of 
initial rotation such that curl free property of superfluid does not get 
violated inside a domain. We assume that only on the circular arc within a
given domain, drawn using the center of the vessel and passing through the 
center of that domain has superfluid velocity as that was of normal component 
before the transition. This will give the gradient of $\theta$ on that arc 
to be the same as given by Eq.(\ref{eq:bias}). 
We can see this by relating velocity of superfluid components 
with normal components on the circular arc, i.e., $v_s=v_n$, which gives 
$|\vec{\nabla} \theta|_{bias} = \frac{m\Omega r}{\hbar}$, which is the same as 
earlier obtained by minimizing the free energy density. It means that 
larger $r$ domain will have more variation in $\theta$ than the domains
with smaller $r$. As we will see, this is precisely the feature that will 
cause the biasing in the formation of vortices over antivortices. 

Now as there is no initial radial flow, we don't expect any radial 
superflow inside a domain also. This  means that $\theta$ will be uniform 
in the radial direction inside each domain. With these 
considerations, we obtain well defined values of 
$\theta$ at every point of a domain. We note that inside a given domain, 
gradient of $\theta$ decreases with increase in $r$, this domain structure 
provides curl free motion of superfluid. So with this, for the rotation of 
the initial normal component whose velocity increases with $r$, after 
becoming superfluid, the velocity becomes $1/r$ dependent inside a given 
domain. This can be viewed as the effect of superfluid transition on the 
velocity profile inside a given correlation domain.
Since with all this, outer domains have stronger variation of $\theta$ (see 
Eq. \ref{eq:bias}), therefore, for the anti-clockwise rotation of vessel, 
we should get more number of vortices than anti-vortices. This bias will depend
upon $\Omega$, system size ($r$ dependence) and also correlation length
$\xi$ (large values of $\xi$ will give larger $\theta$ variation inside 
a domain).  Below we will see that biasing will also depend on the 
inter-domain separation due to modified geodesic rule. 

  We now consider the effect of the bias on the geodesic rule, the way
phase $\theta$ interpolates in between two adjacent domains. 
Conventional Kibble mechanism assumes the {\it geodesic rule} which states
that $\theta$ in between two adjacent domains traces the shortest path
on the vacuum manifold.
Physical motivation for this rule comes from minimizing the free energy 
in the inter-domain region. (As we mentioned, for gauged case, as for
a superconductor, phase variation between two different points is a gauge 
degree of freedom and has no physical significance like gradient energy.
Hence assumption of geodesic rule for gauge case raises conceptual issues,
see ref.\cite{gauge}.) One should note that this {\it conventional} geodesic
rule does not require specification of how large the inter-domain region 
actually is. However, we will see that for the biased case, the physical
extent of the inter-domain region becomes an important parameter.
We will still follow the physical consideration of minimizing the net 
free energy in the inter-domain region. 

For the inter-domain region also we assume that at the center of this region,
the superfluid velocity is the same as the velocity of the initial normal 
fluid component. 
For geodesic rule only the gradient terms of free energy density are 
important, so by ignoring $|\Psi|$ terms from free energy density we have, 

\begin{equation}
 f' = a|\vec{\nabla} \theta|^2 -b|\vec{\nabla} \theta| ,
\end{equation}

\noindent where $a=\frac{\hbar^2}{2m}\Psi_0^2$ and 
$b=\Omega \rho_s r \frac{\hbar}{m}$. We are interested in gradient in the
direction of shortest distance between boundaries of two successive domains.
So in this direction gradient can be written as 
$|\vec{\nabla} \theta| = (\theta_2-\theta_1)/d$, where $\theta_1$ 
and $\theta_2$ are the order parameter values at the boundary of 1st and 2nd
domain respectively when we traverse, in the physical space, from right to 
left (anti-clockwise path) and $d$ is the shortest distance between two 
successive domains. Now we have to determine path for which free 
energy density gets minimized. There are two possible paths on the order 
parameter space. If $\theta_2>\theta_1$, for anti-clockwise path free 
energy density,

\begin{equation}
 f'_1 = a(\theta_2-\theta_1)^2/d^2  -
 b(\theta_2-\theta_1)/d 
 \label{free}
\end{equation}

\noindent and for clockwise path,

\begin{equation}
 f'_2 = a(\theta_2-\theta_1-2\pi)^2/d^2 -
 b(\theta_2-\theta_1-2\pi)/d.  
\end{equation}

\noindent Out of these two paths, one of the path will have lower free energy 
density. Clockwise path will be preferable if condition, $f'_2-f'_1<0$ get 
satisfied, which gives, $\theta_2-\theta_1 > bd/(2a) + \pi$.
Putting values of $a$ and $b$, we get,

\begin{equation}
 (\theta_2-\theta_1) > d |\vec{\nabla} \theta|_{bias} +\pi,
 \label{eq:geod1}
\end{equation}

\noindent which is more restrictive condition to have clockwise path on 
order parameter space than the case when there is no rotation.

Now, if $\theta_2<\theta_1$, free energy density $f'_1$ given by 
Eq.(\ref{free}) will be for clockwise path. For anti-clockwise path 
free energy density will be,

\begin{equation}
 f'_2 = a(\theta_2-\theta_1+2\pi)^2/d^2 -
 b(\theta_2-\theta_1+2\pi)/d.  
\end{equation}
Now in this case, condition $f'_2-f'_1<0$ will be for anti-clockwise 
variation on the order parameter space, which gives,   
\begin{equation}
 \theta_2-\theta_1 < d |\vec{\nabla} \theta|_{bias} - \pi ,
 \label{eq:geod2}
\end{equation}
which is more supportive condition to have anti-clockwise variation of 
$\theta$ than without any rotation. Thus, in both the cases, rotation of 
vessel supports anti-clockwise variation of $\theta$ on the order parameter 
space over clockwise variation even though the path is longer. This shows that
rotation generates biasing in the geodesic rule also.  These modified 
geodesic rules (Eq.(7) and Eq.(9)) will also contribute in the biasing 
of vortices formation over antivortices, along with modified domain structure.  
Note that for Eq.(7) and Eq.(9), we have considered that the variation 
of $\theta$ is along the direction of initial flow. If $\theta$ variation
is considered along a different direction, then suitable projection
of $|\vec{\nabla} \theta|_{bias}$ should be taken.

We consider a cylindrical vessel of radius $R = 40\mu m$, and study the
formation of vortices in an essentially two dimensions system. We have 
taken such a small vessel because of computational limitations. 
Note that effective two dimensions requires that the height of the 
cylinder should be small (i.e. not too large compared to the correlation
length). This will avoid string bending and formation of string loops
which has to be handled in a full three-dimensional simulation. Certainly,
it will be very interesting to see the effects of rotating cylinder in the
formation of strings (including string loops) in a full three-dimensional
simulations and we plan to investigate it in future.  Further, we consider 
correlation length $\xi$ (which determines the domain size) equal to $140 
\mathring{A}$ as an example. This corresponds to a temperature which is just 
below the Ginzburg temperature $T_G$ (as the domain picture is well defined 
only below $T_G$, so defect production is essentially determined just below 
$T_G$, see ref. \cite{kbl}).  For He II system, the critical temperature 
$T_c = 2.17 K$ and Ginzburg temperature $T_G = 2.16 K$ (ref. \cite{tilley}). 
(We mention values of $T_c$ and $T_G$ here ignoring effect of rotation.)
We take inter-domain 
distance $d = 5 \mathring{A}$ (as a sample value, we will discuss
the effect varying $d$ on our results). We have 
considered anti-clockwise rotational of the vessel with angular velocity 
$\Omega \hat{z}$. Critical angular velocity for this system for the
production of vortices using energetics argument, will be $\Omega_{cr}=
\frac{\hbar}{mR^2}\log(R/\xi)\cong 78 ~rad~s^{-1}$ (note that 
radius of the vessel is very small here).

For our two-dimensional simulation, we take a square lattice
with the correlated domains centered at the lattice points. 
Domains are assumed to be circular with radius $\xi$ so that lattice
constant is $(\xi +d)$ with $d$ being the inter-domain separation
as mentioned above. We have performed 
simulation only in first quadrant of the vessel. So the numbers we get
should be multiplied by 4 to get the total number of vortices for the 
whole vessel. Our focus will be
on the probability of vortices per domain. (Note that even for the whole
system, the center of the vessel is within a domain so cannot accommodate
a vortex at that point.) We take the lattice to start from  non-zero 
coordinates (excluding the x and y axes). For winding number calculations
(to locate vortices) we have excluded domains which touch the boundary of
the vessel. 

The essential physics of the Kibble mechanism is implemented 
by taking random $\theta$ value at each lattice points (i.e. at the center 
of domains). We know from the Eq.(\ref{eq:bias}) the gradient of $\theta$ 
at the circular arc, passing through the center of the domain. By knowing
the value of $\theta$ at the center of the domain, and gradient of $\theta$ 
on this arc, we can determine $\theta$ at each point on the arc.
With this, by using the fact that there is no flow in the radial 
direction, so $\theta$ is uniform in this direction, we obtain phase value 
at the domain boundaries which lie on the side of lattice. We also 
use modified geodesic rule Eq.(\ref{eq:geod1}) and Eq.(\ref{eq:geod2}) for 
variation of $\theta$ in the inter-domain region. To implement this rule, as 
we mentioned, we assume that at the center-point of inter-domain 
region (which is the middle point of a link) superfluid has same velocity 
as was of normal components before the transition (given by 
Eq.(\ref{eq:bias})). We project this velocity along the direction of 
lattice side to get $\vec{\nabla} \theta$ along the lattice side.
With this, and knowing the values of $\theta$ at domain boundaries, we
implement the modified geodesic rule Eq.(\ref{eq:geod1}) and 
Eq.(\ref{eq:geod2}) to know $\theta$ variation in that region. 
With all this, we calculate winding in each plaquette. Depending upon 
the winding, at the center of plaquette we obtain vortices or antivortices.

\begin{figure}
 \includegraphics[width=0.85\linewidth]{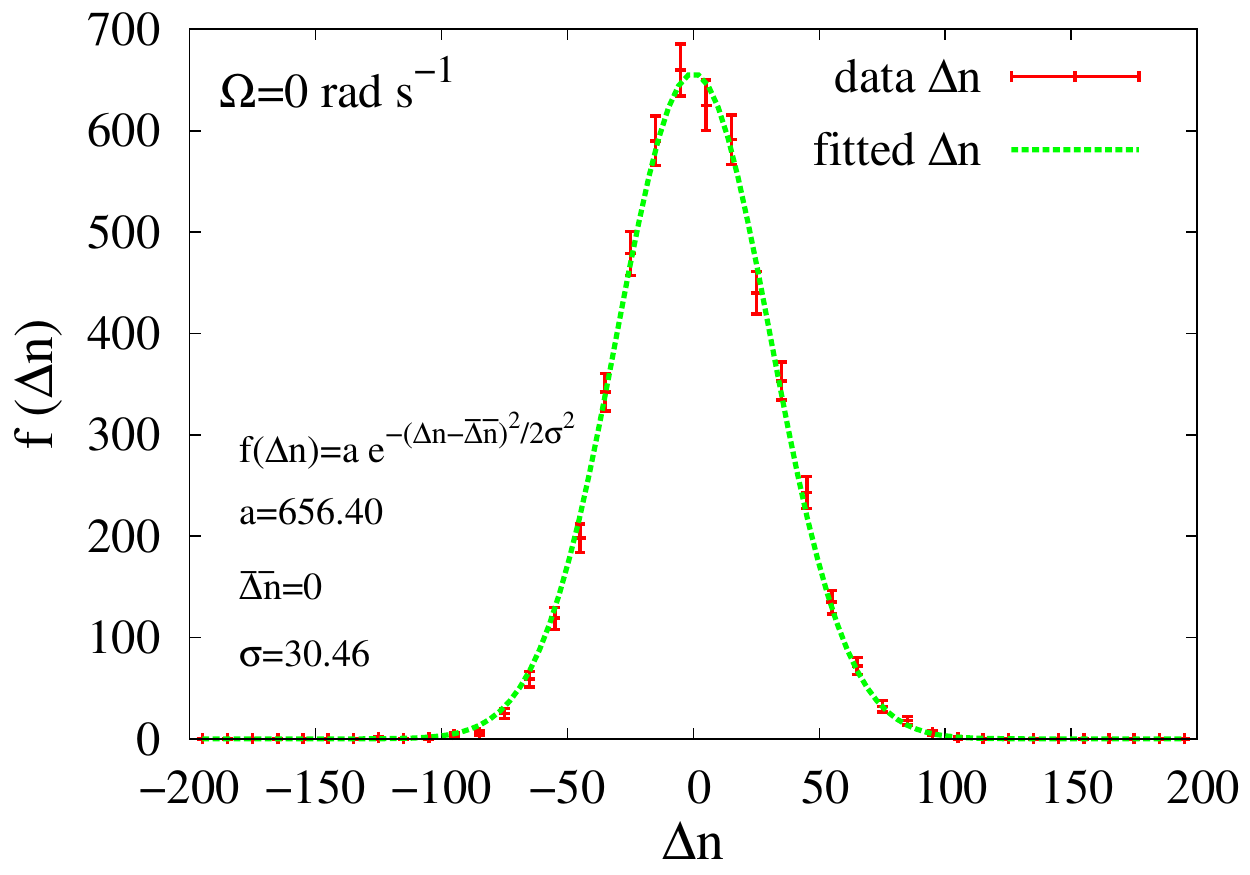}
 \includegraphics[width=0.85\linewidth]{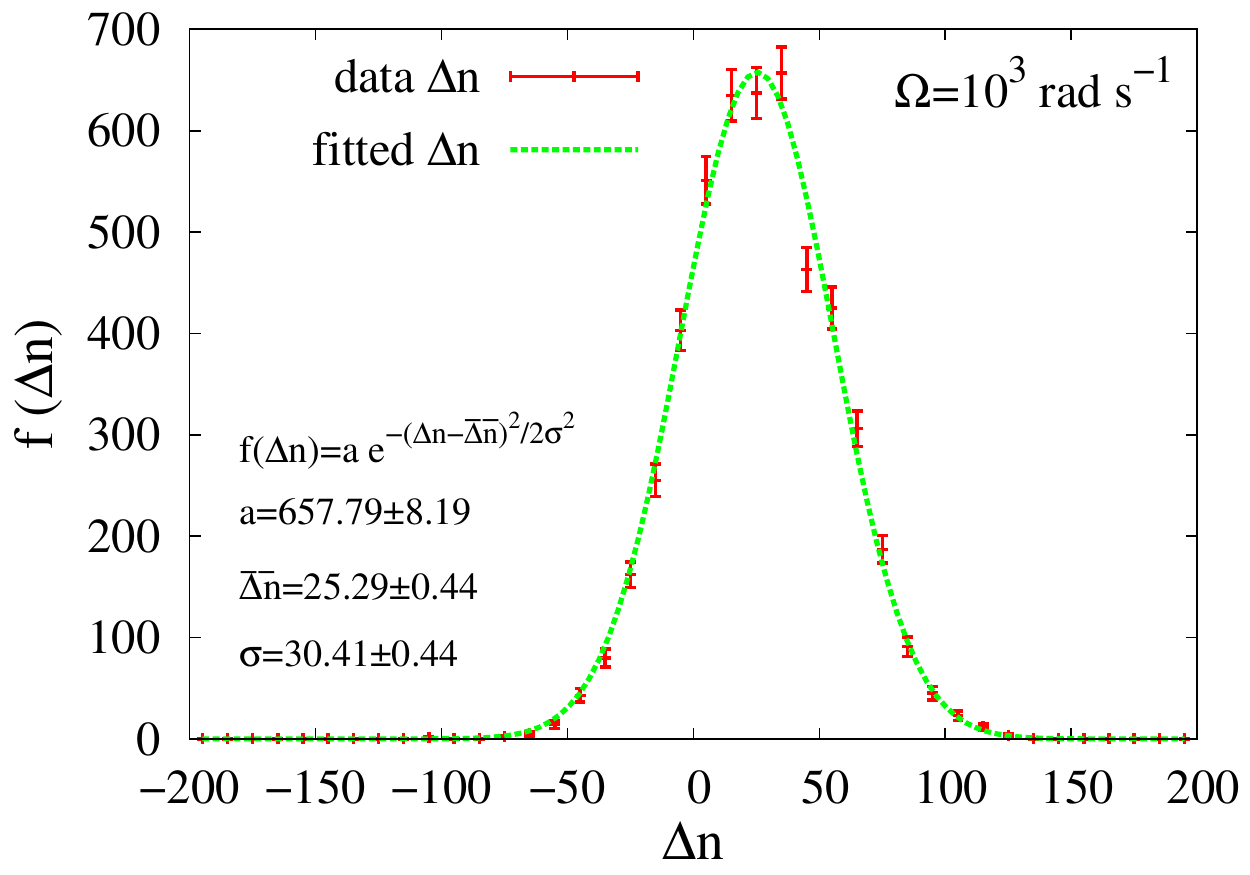}
\caption{Distribution of vortices $-$ antivortices. Upper plot shows
the case without any rotation of the vessel ($\Omega = 0$) giving
the mean value of Gaussian distribution, $\overline{\Delta n}=0$. Lower
plot corresponds to the case with angular
velocity of the vessel $\Omega = 10^3$ $rad$ $s^{-1}$ showing that
$\overline{\Delta n}$ gets shifted from zero to value 25, showing 
a net biasing in formation of vortices over antivortices. 
As we simulate only a quadrant, the full vessel will
give value of net $\overline{\Delta n}$ of about 100.} 
\label{fig1}
 \end{figure}

Now we present the results of our simulation.  We consider different values of
the angular velocity $\Omega$, and for each $\Omega$ we generate 5000
events for defect formation to get good statistics of vortex-anti-vortex 
production. Fig.\ref{fig1} shows the distribution of net defect number
$\Delta n$ (= defect number $-$ anti-defect number) for 5000 events.  
Upper plot shows the distribution without any rotation of vessel 
($\Omega = 0$), we get standard distribution as predicted by the Kibble 
mechanism. This distribution follows Gaussian distribution $f(\Delta n)=
a e^{-\frac{(\Delta n-\overline{\Delta n})^2} {2\sigma^2}}$. By fitting 
the distribution, we obtain the  parameters of this Gaussian as: $a=656.40, 
~\overline{\Delta n}\cong 0,~\sigma=30.46$ (we have taken bin width 10
with error bars on the plot taken as $[f(\Delta n)]^{1/2}$ for each bin value). 
Important point to note is that center of Gaussian $\overline{\Delta n}$ has 
zero value which is the standard prediction of Kibble mechanism; no biasing 
in the formation of vortices and antivortices (on the average). We obtained 
average total number of defects from the simulation to be $ N = 1857948$. 
Kibble mechanism makes an important prediction of relation between $\sigma$ 
and $N$ Ref. \cite{lc3}, $\sigma=CN^{\nu}$, where value of $C$ for square 
domains is 0.71. The exponent $\nu$ is universal and its theoretical value is 
$\nu=1/4$ for the present case.  From the obtained value of $\sigma$ 
and $N$ with simulation, we derive value of $\nu=0.2604$, which is 
quite close to the theoretical value $0.25$ and matches well with the
experimental value of $\nu = 0.26 \pm 0.11 $ obtained for liquid crystal 
case, see ref.\cite{lc3}. 

The lower plot in Fig.1 gives the distribution of $\Delta n$ for the case of
vortex formation during superfluid transition in a rotating vessel with angular 
velocity $10^3~rad~s^{-1}$. We see that in this case also we get a Gaussian 
distribution but shifted with the mean value $\overline{\Delta n}=25$, which 
clearly shows that there is a biasing in the formation of vortices over 
antivortices. For the whole cylinder, we thus expect to get on an 
average more than 100 vortices over antivortices in the vessel. This
bias in the net value of $\Delta n$  occurs here because of the modification 
in the domain structure and geodesic rule in the presence of rotation. 
Thus our proposed modification of the Kibble mechanism, with modified
domain structure along with the modified geodesic rule, is able to
accommodate the  expected bias in the net value of $\Delta n$ due
to the rotation of the vessel. 

  Table I shows the obtained values of $\overline{\Delta n}$, $\sigma$, and 
$N$ from simulations at different $\Omega$ values; for each $\Omega$ we 
have performed $5000$ events. Values of $\nu$ is obtained from the relation 
$\sigma=CN^{\nu}$, $C=0.71$.  It is very clear that with $\Omega$
all the other quantities are increasing. 
 
\begin{table}[ht]
\caption{Effect of rotation on the formation of vortices} 
\centering 
\begin{tabular}{c c c c c} 
$\Omega$ & $\overline{\Delta n}$ & $\sigma$ & $N$ & $\nu$ \\ [0.5ex] 
\hline 
0 & 0.0 & 30.46 &~ 1857948 &~0.2604 \\ 
$10^3$ &~ 25.29$\pm$0.44 &~ 30.41$\pm$0.44 &~ 1858005 &~ 0.26029 \\
$10^4$ &~ 250.20$\pm$0.26 &~ 30.90$\pm$0.26 &~ 1858003 &~ 0.2614 \\
$10^5$ &~ 2492.88$\pm$0.30 &~ 31.42$\pm$0.30 &~ 1858010 &~ 0.26255 \\
$5\times10^5$ &~ 12466.9$\pm$0.36 &~ 31.77$\pm$0.35 &~ 1858031 &~ 0.26332 \\
$10^6$ &~ 24932.8$\pm$0.34 &~ 35.81$\pm$0.32 &~ 1858136 &~ 0.27161 \\
$10^7$ &~ 240603.$\pm$4.96 &~ 43.47$\pm$0.35 &~ 2745682 &~ 0.27753 \\ [1ex] 
\hline 
\end{tabular}
\label{table:nonlin} 
\end{table}

\begin{figure}
 \includegraphics[width=0.85\linewidth]{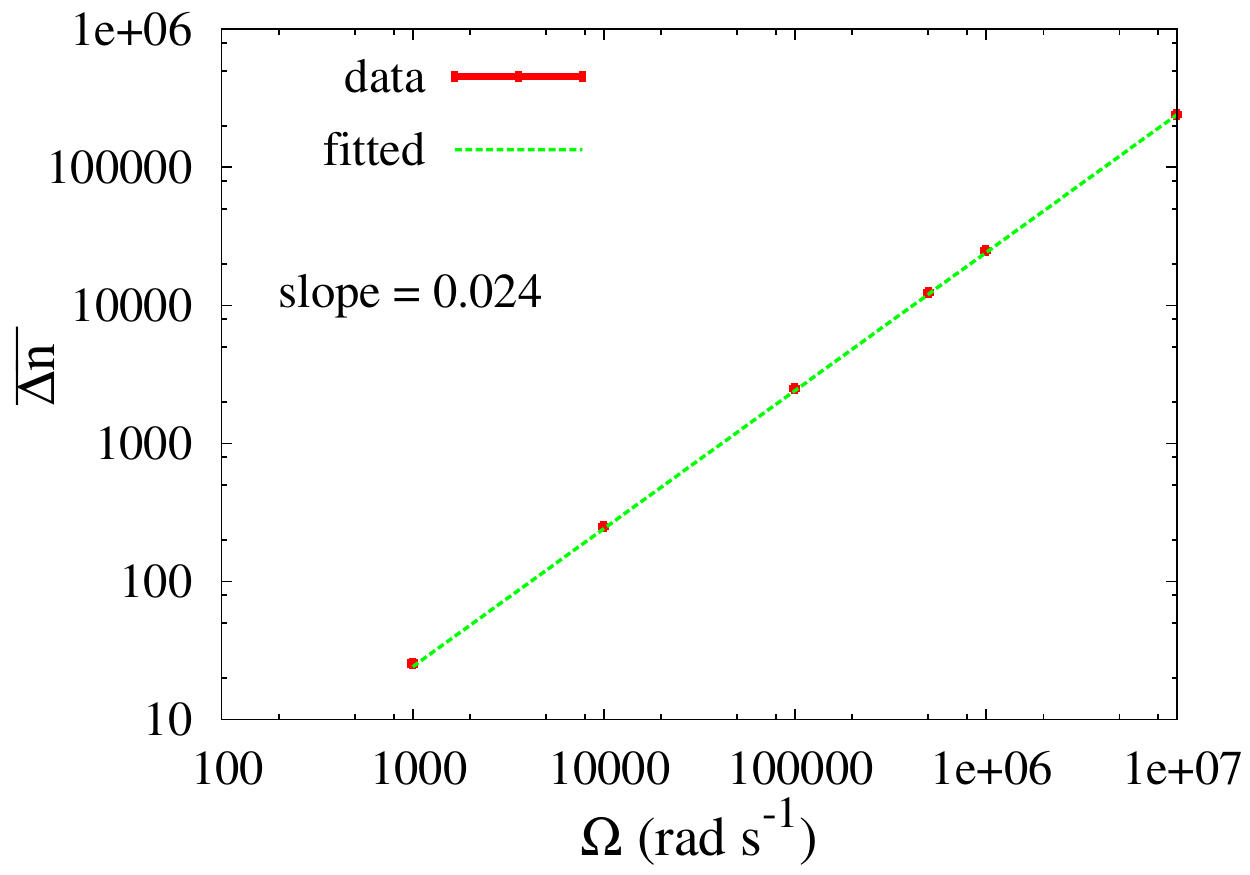}
 \caption{Variation of $\overline{\Delta n}$ with $\Omega$ in log-log scale.
 This plot shows that $\overline{\Delta n}$ linearly depends on $\Omega$
 with slope 0.024. Slope will be about 0.1 if simulation perform in
 full cylinder.}
 \label{fig2}
\end{figure}

Fig.\ref{fig2} shows the dependence of $\overline{\Delta n}$ on $\Omega$
(axes are in log-log scale).  This plot clearly shows that
$\overline{\Delta n}$ linearly increases with $\Omega$ with slope $0.024$. 
Slope will be about 0.1 (4 times higher) for the full cylindrical vessel.  
As shown in Table I, for $\Omega = 0$ we find $\overline{\Delta n} = 0.0$
as expected from the usual Kibble mechanism. However, the straight line fit
in Fig.1 does not pass through the origin (0,0) of the plot, instead
it gives $\overline{\Delta n} \simeq 1.0$ for $\Omega = 0$.  The best fit 
line is given by $\overline{\Delta n} = 0.024 \Omega + 1.0$. For full
vessel this would mean $\overline{\Delta n} \simeq 4$ at $\Omega = 0$. 
This is clearly due
to fluctuations in the simulation results for finite number of runs.
With the plot in Fig.2, at the critical angular velocity $\Omega_{cr}$  
($\simeq 78$ rad s$^{-1}$ as mentioned earlier) we will have on an average 
net 12 vortices (for the whole vessel). Note when number of vortices is
calculated using only energetics arguments in the vortex model, we expect
a single vortex at the critical angular velocity. 
However, just after the superfluid
transition, number of vortices also gets contributions from the Kibble
mechanism (suitably modified as proposed here) whose contributions have
a Gaussian spread with $\sigma$ as given in Table 1. Thus the final
value of $\overline{\Delta n}$ will be expected to deviate from the
vortex model prediction in general.
 It is still interesting to ask that with proper incorporation of the
Kibble vortices, what is the {\it new} critical angular velocity at which
one expects to get $\overline{\Delta n} = 1$. With our results,
angular velocity of the vessel will be smaller than a different critical 
velocity, say, $\Omega_{Kibble}$, which also depends on system parameters 
system size, the inter-domain separation $d$, etc. 
It is very interesting to study the behavior of 
$\Omega_{Kibble}$ in comparison to $\Omega_{cr}$ and we plan to study this
in future. Especially interesting will be to investigate the dependence of 
our results on the parameter $d$. For a first order transition, with
a simple situation of nucleation of a large density of critical bubbles
(almost at close packing) the value of $d$ will be given by $2\times$ the
bubble wall thickness (while $\xi$ corresponds to the bubble diameter).
By considering different experimental situations, the ratio $d/\xi$ can
be varied and its effects on various results, especially on $\Omega_{Kibble}$
can be studied. For a second order transition such a study will be
more complicated. In view of these issues, it is clear that a proper 
interpretation of Hess and Fairbank experiment \cite{hess2} requires a 
more detailed analysis. Measurement of average number of vortices in 
experiment with sufficiently large number of events for  superfluid 
transition with angular velocity just below $\Omega_{cr}$ 
may give a good test for the model here we propose. A non-zero value 
of angular momentum of superfluid below $\Omega_{cr}$ will give a solid 
support for this model. It will also show that there is a critical 
angular velocity $\Omega_{kibble}$ which is different from  
$\Omega_{cr}$ for the phase transition in the presence of rotation.

 As mentioned above, the best fit line for results in Table I
gives $\overline{\Delta n} = 0.1\Omega$ (ignoring the intercept,
hence for large $\Omega$). This  matches very well with the vortex model 
prediction which gives $n \simeq 2\pi R^2 m\Omega/h \simeq 0.1 \Omega$ 
(Ref. \cite{tilley}). This is expected as for very large $\Omega$, number 
of vortices should be dominated by the effects of rotation.
We again mention that our results depend on various parameters, such as
$\xi, d$ etc. Thus one needs to study whether this agreement with
the vortex model prediction (for large $\Omega$) is valid in general.

We emphasize that the free energy of individual defects plays no role in 
the Kibble mechanism (even with the modifications we propose). Still, with
our incorporation of initial rotation of the normal fluid (and its some 
fraction getting transferred to the superfluid flow after the transition)
at least some part, if not all, of the "rotation induced vortices" have 
been included in this proposed modified Kibble mechanism. This point
will be particularly important for small rotations where very few vortices
are expected from energetics arguments. This modified Kibble mechanism
gives defect density right after the transition which will evolve in time,
and approach the density expected using equilibrium free energy
arguments.  Thus, if the (modified) Kibble mechanism gives lesser number of net 
produced vortices then with time, more number of vortices 
will get produced and ultimately in the equilibrium, system will have $n$
number of vortices as predicted by the vortex model using energetics arguments.
It is also interesting to study the distribution of vortices
and antivortices as a function of distance from center in our model. The
equilibrium distribution is uniform but as mentioned above, the distribution
right after the transition may be different due to non-equilibrium
contributions from the (modified) Kibble mechanism. A non-uniform initial
distribution will have very important implications for the case of neutron stars
where migration of vortices to achieve uniform (equilibrium) distribution
will lead to change in moment of inertia of the neutron star (as in the model
discussed in \cite{pulsar}).  This  requires large
statistics and this study is underway. 

Table I also shows that the width of the Gaussian $\sigma$ increases
with $\Omega$ (slowly initially but strongly for large values of $\Omega$).
$\sigma$ represent randomness in the formation of vortices and anti-vortices.
If formation of vortices and antivortices is completely uncorrelated then value 
of $\sigma$  goes like $\sim N^{1/2}$; width of Binomial distribution. But 
there is correlation between production of defect and anti-defects in 
Kibble mechanism (Ref.\cite{lc3}) causing suppression in randomness and 
hence $\sigma \sim N^{1/4}$. By writing $\sigma \sim N^\nu$ we see from
the Table I, that $\nu$ increases with $\Omega$ showing that correlation 
between production of vortices and antivortices is getting suppressed with 
$\Omega$. We also fit the dependence of $\sigma$ on $\Omega$. 
A reasonable fit for $\sigma$ as a function of $\Omega$ is obtained by 
$\sigma = a\Omega^p +b$ where fitted values of parameters are found to be
$a = 0.004 \pm 0.006, p = 0.51 \pm 0.10, b = 30.30 \pm 0.65$.
Even though value of $a$ is entirely dominated by error, this fit does 
suggest a systematic variation of $\sigma$ with $\Omega$ with exponent
$p \simeq 0.5$. We plan to carry out a systematic study of this result
and increase of $\nu$ with $\Omega$ in future.

\begin{figure}
 \includegraphics[width=0.85\linewidth]{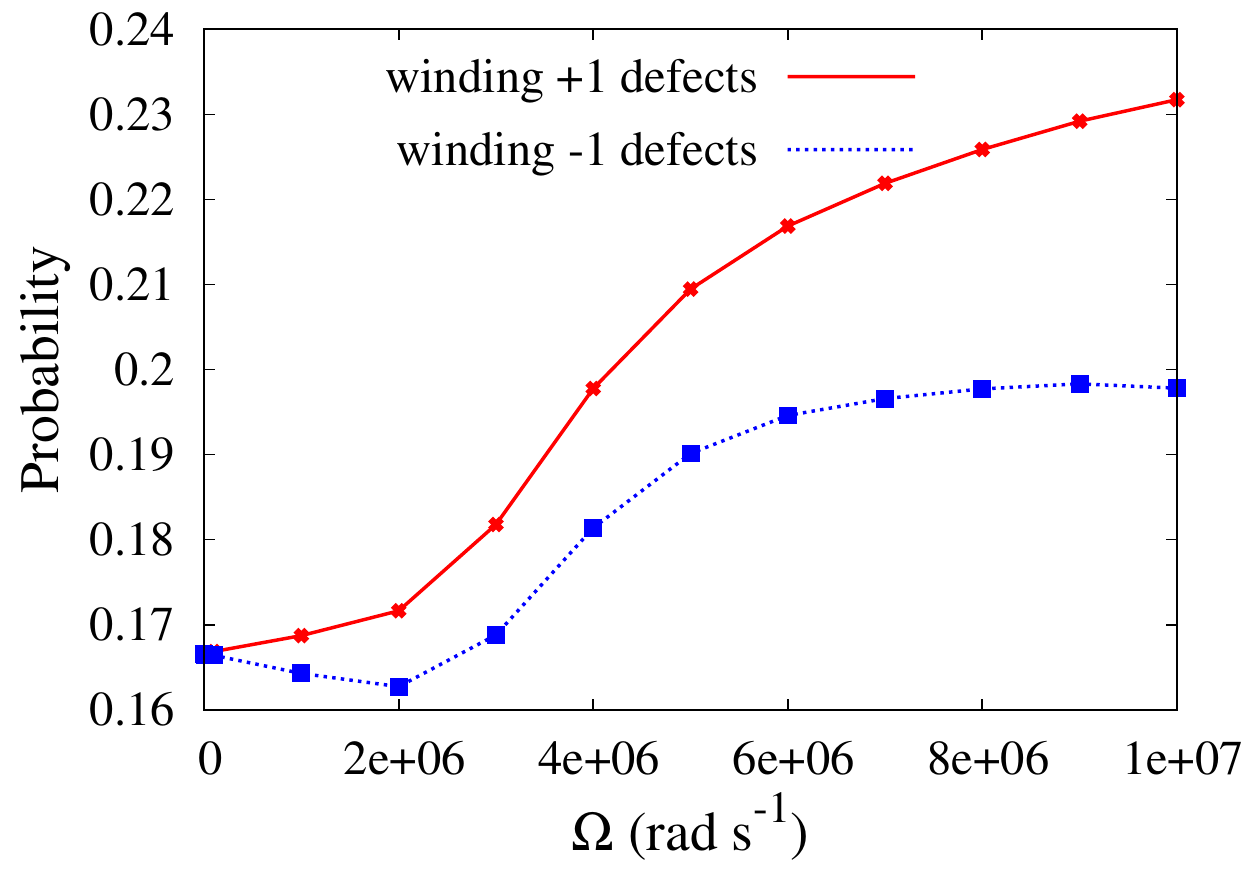}
 \caption{Plot shows probability of formation of single winding defects 
 and anti-defects as a function of $\Omega$. Probabilities for both the 
 cases changes differently with $\Omega$ and causing biasing 
 in the formation of defects over anti-defects.}
 \label{fig5}
\end{figure}

Fig.\ref{fig5} presents results for a single event for the
number of defects per domain, i.e., probability of formation of defects.
Fig.\ref{fig5} shows probability of formation of single winding defects 
and anti-defects as a function of $\Omega$.  We note that both probabilities
increase with $\Omega$, with winding $+1$ defect probability increasing
faster than the probability for winding $-1$ (anti-defects), reflecting 
biasing in the formation of defects over anti-defects. The total defect 
number (defects + anti-defects)  probability
increases  with $\Omega$ as expected.

We also find an increase in the formation of winding number two defects 
and anti-defects as a function of $\Omega$ (we have not included
those numbers here). Probabilities for both the 
cases become non-zero at $\Omega>2\times 10^6$ $rad$ $s^{-1}$ and changes 
differently with $\Omega$, again reflecting biasing in the formation 
of defects over anti-defects. It is well known fact that winding number 
two defects are unstable in superfluid systems and split into two single 
winding defects eventually enhancing single winding defects formation 
probabilities. We have also checked the effects of varying the inter-domain
separation $d$ on our results. For $\Omega = 10^6$, increase of $d$ by a 
factor of 20 (from $d = 1 \mathring{A}$ to $d = 20 \mathring{A}$) 
increases probabilities for winding one defect as well
as antidefect by about 15 \%. Change in winding two defect probabilities
is very small and dominated by fluctuations. For smaller $\Omega = 10^5$
the change in probabilities is very small and dominated by fluctuations.
The effect of $d$ on various probabilities is a complex issue and we plan to
study it systematically in future. We note that while increase of vortex
formation probability is expected as a function of increasing angular velocity, 
it may appear puzzling why anti-defect probability also increases with
the rotation. The explanation for this may lie in the correlation of defects
and antidefects which is an important and non-trivial prediction of
the Kibble mechanism. As we see from Table I, the defect-antidefect 
correlation exponent $\nu$, while increasing slightly with angular velocity
to a value of about 0.28, still remains far below the value of 0.5
for uncorrelated case. Thus, while vortex probability increases naturally
with the rotation, the underlying domain structure forces larger 
probability of formation of anti-vortices close to vortices for
winding number 1 as well as for winding number 2 case. (Basically
from the fact that positive winding across two domains appears as
anti-winding for the neighboring region.) 

 Experimental tests of our predictions based on this modified Kibble 
mechanism will lend strong support to the whole underlying picture
of the Kibble mechanism which is adaptable for varying experimental
conditions such as biased formation of flux tubes in superconductors in the
presence of external field etc.  We mention here an important aspect
of vortex formation in superfluids via the Kibble mechanism which is
not present for other types of topological defects (as emphasized in our 
earlier work \cite{cfl}). We mentioned above that we assume that part of 
normal component which undergoes superfluid condensation carries the same 
angular momentum as it had before the transition (along an arc at the
center of the domain). This just reflects the local conservation of linear 
momentum during the superfluid transition on that arc.
However, even if there was no initial motion of the fluid, still
during phase transition, spontaneous generation of flow of the superfluid 
will arise simply from the spatial variation of the condensate phase.
Indeed, it is this (random) phase variation from one domain to another
which leads to formation of vortex network and hence spontaneous
generation of superflow. What happens then to local linear momentum 
conservation? Basically, some fraction of ($^4$He) atoms form the
superfluid condensate during the transition and develop momentum
due to the non-zero gradient of the phase of the condensate. The only
possibility is that the remaining fraction of atoms (which form the
normal component of fluid in the two-fluid picture) develop opposite
linear momentum so that the momentum is locally conserved. (Here we avoid
conceptual question of an ideal instantaneous quench to almost zero 
temperature where there is no normal component left). This means
that there is no net momentum flow anywhere right after the transition.
For superfluid transition in a rotating vessel, same consideration will 
apply to the normal component in a domain
in regions away from the central arc as in those regions superflow will
in general not match with the initial flow due to rotation implying
generation of extra counterbalancing normal flow component.
Note, this argument is quite different from the conventional
argument of net angular momentum conservation for Kibble superfluid
vortices where one knows that spontaneous generation of net rotation of
the superfluid has to be counter balanced by the opposite rotation of the
vessel containing the superfluid \cite{zrk1}. Here, we are arguing for local 
linear momentum conservation which implies generation of complex flow
pattern for normal component depending on the generation of spontaneous
part of the superflow during the transition.  The final picture is then that, 
the original
rotation of the normal fluid (before the transition) is simply transferred 
to the rotation of the superfluid which, via our modified Kibble mechanism,
accounts for the net bias of vortices over anti-vortices. At the same time    
generation of extra vortices and anti-vortices via the random domain
formation (via the Kibble mechanism) leads to extra local superfluid 
circulation in the system which will be accompanied by opposite 
circulation being generated in the normal component of the fluid (to 
balance the momentum conservation). To incorporate both these contributions
accurately, one must carry out simulations of the transition with a two
fluid picture in a rotating vessel. These consideration must be incorporated 
for any experimental test of the Kibble mechanism (either the conventional 
one, or the modified one presented here). It is possible that a due
consideration of  this spontaneously generated counterbalancing flow of the
normal fluid may improve agreement of the results of various superfluid helium
experiments with the Kibble mechanism.  We plan to carry
out a detailed investigation of this issue in a future work.

  In conclusion, we have proposed a modification of the conventional
Kibble mechanism for the situation of   production of
topological defects when physical situation requires excess of 
windings of one sign over the opposite ones. We have considered the case
of formation of vortices for superfluid $^4He$ system when the transition
is carried out in a rotating vessel. As our results show, this biased formation 
of defects can strongly affect the estimates of net defect density. Also, 
these studies may be crucial in discussing the predictions relating 
to defect-anti-defect correlations. The modified Kibble mechanism
we presented here has very specific predictions about net defect number
which shows a clear pattern of larger fluctuations (about mean value
governed by the net rotation) compared to the conventional Kibble prediction. 
This can be easily tested in experiments. Further, even the average net
defect number deviates from the number obtained from energetics considerations,
especially for low values of $\Omega$.
This implies that exactly at the time of transition, a different net defect
number will be formed on the average, which will slowly evolve to a value
obtained from energetic considerations. These considerations can be extended 
for the case of flux tube formation in superconductors (with appropriate 
modifications for the gauged case), and we hope to present it in a future work.
Such a modified Kibble mechanism is also needed to study formation of 
baryons at finite chemical potential in the framework of chiral sigma model
where baryons appear as Skyrmions which are topological solitons (extending
our earlier work on 1+1 D Skyrmion formation to 3+1 D \cite{skyrmi}). 
Our results will have implications for superfluid transition in
rotating neutron stars (where phase transition induced density fluctuations
could be detected by observing pulsar signal changes, as proposed
by some of us \cite{pulsar}). In an earlier work \cite{cfl}, we 
considered the possibility of superfluid phases of QCD, e.g. neutron 
superfluid and color-flavor-locked phase, in low energy heavy-ion collisions
and showed that this will lead to production of few vortices via the
(conventional) Kibble mechanism which can 
strongly affect the hydrodynamical evolution of the system and can be 
detected by measuring flow fluctuations. For low energy non-central 
collisions superfluid phase transition is likely to happen in the presence 
of an overall rotation of the plasma region. Resulting vortex production
for such a case must be studied by a modified Kibble mechanism, as
we have proposed here.

\begin{acknowledgements}
We are very grateful to Sanatan Digal for careful reading of manuscript
and giving valuable suggestions, especially for pointing out that
increased antivortex probability may relate to defect-antidefect
correlation.  We thank Srikumar Sengupta, P.S. Saumia, Abhishek Atreya, 
Oindrila Ganguly, Partha Bagchi, Arpan Das, Minati Biswal, and Priyo 
Shankar Pal for very useful suggestions and comments. 
\end{acknowledgements}


\begin{thebibliography}{99}

\bibitem{kbl} T.W.B. Kibble, J. Phys. {\bf A9}, 1387 (1976);
Phys. Rept. {\bf 67}, 183 (1980).

\bibitem{zrk1} W.H. Zurek, Nature {\bf 317}, 505 (1985).

\bibitem{zrk2} W.H. Zurek, Phys. Rep. {\bf 276}, 177 (1996).  

\bibitem{rjnt} A. Rajantie, Int. J. Mod. Phys. {\bf A17}, 1 (2002). 

\bibitem{he} P.C. Hendry et al, Nature (London) {\bf 368}, 315 (1994);
V.M.H. Ruutu et al, Nature (London) {\bf 382}, 334 (1996); M.E. Dodd
et al, Phys. Rev. Lett. {\bf 81}, 3703 (1998); R. Carmi et al,
Phys. Rev. Lett. {\bf 84}, 4966 (2000); see also, G.E. Volovik,
Czech. J. Phys. {\bf 46}, 3048 (1996).

\bibitem{sccor} R. Carmi, E. Polturak, and G. Koren, Phys. Rev.
Lett. {\bf 84}, 4966 (2000); A. Maniv, E. Polturak, and G. Koren,
cond-mat/0304359; R.J. Rivers and A. Swarup, cond-mat/0312082;
E. Kavoussanaki, R. Monaco, and R.J. Rivers, Phys. Rev. Lett.
{\bf 85}, 3452 (2000); S. Rudaz, A.M. Srivastava and S. Varma,
Int. J. Mod. Phys. {\bf A14}, 1591 (1999)1.

\bibitem{lc1} I. Chuang, R. Durrer, N. Turok and B. Yurke,
Science 251, 1336 (1991); R. Snyder et al.
Phys. Rev. {\bf A45}, R2169 (1992); I. Chuang et al.
Phys. Rev. {\bf E47}, 3343 (1993).

\bibitem{lc2} M.J. Bowick, L. Chandar, E.A. Schiff and A.M. Srivastava,
Science 263, 943 (1994).

\bibitem{lc3}  S. Digal, R. Ray, and A.M. Srivastava, Phys. Rev.
Lett. {\bf 83}, 5030 (1999); R. Ray and A.M. Srivastava, hep-ph/0110165.

\bibitem{landau} L.D. Landau and E.F. Lifshitz, {\bf Statistical Physics 
Part 2}, Pergamon Press Ltd. (1980).

\bibitem{hess1} G.B. Hess, Phys. Rev. {\bf 161}, 189 (1967).

\bibitem{gauge} S. Rudaz and A.M. Srivastava, Mod. Phys. Lett. 
A8, 1443 (1993).

\bibitem{hess2} G.B. Hess and W.M. Fairbank, Phys. Rev. Lett. {\bf 19}, 
216 (1967).

\bibitem{tilley} D.R. Tilley and J. Tilley, {\bf Superfluidity and 
Superconductivity}, Third Edition, Overseas Press (2005).

\bibitem{skyrmi} V.S. Kumar, B. Layek, A.M. Srivastava, S. Sanyal, 
and V.K. Tiwari, IJMP A {\bf 21}, 1199-1219 (2006).

\bibitem{pulsar} P. Bagchi, A. Das, B. Layek, A. M. Srivastava, 
Phys.Lett. {\bf B747}, 120-124 (2015).

\bibitem{cfl} A. Das, S.S. Dave, S. De, A.M. Srivastava, Mod. Phys. 
Lett. A {\bf 32}, 1750170 (2017).

\end{thebibliography}
\end{document}